\documentclass[twocolumn]{article}
\usepackage[round]{natbib}
\usepackage{times}
\usepackage{latexsym}
\ifx\pdfoutput\undefined
\usepackage{graphicx}
\else
\usepackage[pdftex]{graphicx}
\usepackage{epstopdf}
\fi
\usepackage{hyperref}
\usepackage{xcolor} 

\title{Universal Masking is Urgent in the COVID-19 Pandemic: 
  \\  SEIR and Agent Based Models, Empirical Validation, 
  \\  Policy Recommendations 
}
\date{21 April 2020}
\author{
  \textbf{De Kai} \small{\textsc{PhD MBA}}\\\vspace{-0.3em}
  \small{HKUST (University of Science \& Technology), Hong Kong}\\\vspace{-0.3em}
  \small{International Computer Science Institute, Berkeley, CA, USA}\\\vspace{-0.3em}
  \small{dekai@cs.ust.hk}\ \ \small{dekai@icsi.berkeley.edu}\\\vspace{-0.3em}
  \small{@dekai123}\ \ \small{http://dek.ai}
  \and
  \textbf{Guy-Philippe Goldstein} \small{\textsc{MBA}}\\\vspace{-0.3em}
  \small{Ecole de Guerre Economique, Paris, France}\\\vspace{-0.3em}
  \small{guyphilippeg@gmail.com}\\\vspace{-0.3em}
  \small{@guypgoldstein}
  \and
  \textbf{Alexey Morgunov}\\\vspace{-0.3em}
  \small{University of Cambridge, UK}\\\vspace{-0.3em}
  \small{Manifold Research, Cambridge, UK}\\\vspace{-0.3em}
  \small{asm63@cam.ac.uk}\ \ \small{alexey@manifoldresearch.com}\\\vspace{-0.3em}
  \small{@AlexeyMorgunov}
  \and
  \textbf{Vishal Nangalia} \small{\textsc{PhD MBChB FRCA}}\\\vspace{-0.3em}
  \small{University College London, UK}\\\vspace{-0.3em}
  \small{ELU AI Ltd, London, UK}\\\vspace{-0.3em}
  \small{Royal Free Hospital, London, UK}\\\vspace{-0.3em}
  \small{vishal.nangalia@gmail.com}\\\vspace{-0.3em}
  \small{@v\_alien}
  \and
  \textbf{Anna Rotkirch} \small{\textsc{PhD}}\\\vspace{-0.3em}
  \small{Population Research Institute, The Family Federation of Finland}\\\vspace{-0.3em}
  \small{anna.rotkirch@vaestoliitto.fi}\\\vspace{-0.3em}
  \small{@AnnaRotkirch}\ \ \small{https://blogs.helsiki.fi/rotkirch}
}
\date{21 April 2020}
\begin{document}
\maketitle

\newcommand\blfootnote[1]{%
  \begingroup
  \renewcommand\thefootnote{}\footnote{#1}%
  \addtocounter{footnote}{-1}%
  \endgroup
}
\begin{abstract} 
\par We present two models for the COVID-19 pandemic predicting the impact of universal face mask wearing upon the spread of the SARS-CoV-2 virus—one employing a stochastic dynamic network based compartmental SEIR (susceptible-exposed-infectious-recovered) approach, and the other employing individual ABM (agent-based modelling) Monte Carlo simulation—indicating (1) significant impact under (near) universal masking when at least 80\% of a population is wearing masks, versus minimal impact when only 50\% or less of the population is wearing masks, and (2) significant impact when universal masking is adopted early, by Day 50 of a regional outbreak, versus minimal impact when universal masking is adopted late. These effects hold even at the lower filtering rates of homemade masks. To validate these theoretical models, we compare their predictions against a new empirical data set we have collected that includes whether regions have universal masking cultures or policies, their daily case growth rates, and their percentage reduction from peak daily case growth rates. Results show a near perfect correlation between early universal masking and successful suppression of daily case growth rates and/or reduction from peak daily case growth rates, as predicted by our theoretical simulations. 
  \blfootnote{*This collective work grew out of a Kinnernet discussion group about COVID-19 initiated by Guy-Philippe Goldstein. All authors contributed to the overall design and writing. Additionally, Goldstein formulated overall study goals and analysed policy data, Morgunov ran the SEIR simulation and collected policy data, De Kai created the online interactive ABM simulation, Nangalia contributed with medical expertise and to the model design, and Rotkirch and De Kai first drafted the report.}

\par Taken in tandem, our theoretical models and empirical results argue for urgent implementation of universal masking in regions that have not yet adopted it as policy or as a broad cultural norm. As governments plan how to exit societal lockdowns, universal masking is emerging as one of the key NPIs (non-pharmaceutical interventions) for containing or slowing the spread of the pandemic. Combined with other NPIs including social distancing and mass contact tracing, a ``mouth-and-nose lockdown'' is far more sustainable than a ``full body lockdown'', from economic, social, and mental health standpoints. To provide both policy makers and the public with a more concrete feel for how masks impact the dynamics of virus spread, we are making an interactive visualization of the ABM simulation available online at http://dek.ai/masks4all. We recommend immediate mask wearing recommendations, official guidelines for correct use, and awareness campaigns to shift masking mindsets away from pure self-protection, towards aspirational goals of responsibly protecting one's community.\end{abstract} 
\section{Introduction}
\par With almost all of the world's countries having imposed measures of social distancing and restrictions on movement in March 2020 to combat the COVID-19 pandemic,~governments now seek a sustainable pathway back towards eased social restrictions and a functioning economy. Mass testing for infection and serological tests for immunity, combined with mass contact tracing, quarantine of infected individuals, and social distancing, are recommended by the WHO and have become widely acknowledged means of controlling spread of the SARS-CoV-2 virus until a vaccine is available.
\par Against this backdrop, a growing number of voices suggest that universal face mask wearing, as practiced effectively in most East Asian regions, is an additional, essential component in the mitigation toolkit for a sustainable exit from harsh lockdowns. The masks-for-all argument claims that~``test, trace, isolate'' should be expanded to~``test, trace, isolate, mask''. This paper presents cross-disciplinary, multi-perspective arguments for the urgency of universal masking, via both new theoretical models and new empirical data analyses. Specifically, we aim to illustrate how different degrees of mass face wearing affects infection rates, and why the timing of introduction of universal masking is crucial.
\par In the first of two new theoretical models, we introduce an SEIR~ (susceptible-exposed-infectious-recovered) model of the effects of mass face mask wearing over time compared to effects of social distancing and lockdown.~In the second of two new theoretical models, we introduce a new interactive individual ABM (agent-based modelling) Monte Carlo simulation showing how masking significantly lowers rates of transmission. Both models predict significant reduction in the daily growth of infections on average under universal masking (80-90\% of the population) if instituted by day 50 of an outbreak, but not if only 50\% of the population wear masks or if institution of universal masking is delayed.
\par We then compare the two new simulations presented here against a new empirical data set we have collected that includes whether regions have universal masking cultures or policies, their daily case growth rates, and their percentage reduction from peak daily case growth rates. Since little precise quantitative data is available on cultures where masking is prevalent, we explain in some depth the historical and~sociological factors that support our classification of masking cultures. Results show a near perfect correlation between early universal masking and successful suppression of daily case growth rates and/or reduction from peak daily case growth rates, as predicted by our theoretical simulations.
\par To preview the key policy recommendations that our two new SEIR and ABM predictive models and empirical validation all lead to:
\begin{enumerate}
\item Masking should be mandatory or strongly recommended for the general public when in public transport and public spaces, for the duration of the pandemic.
\item Masking should be mandatory for individuals in essential functions (health care workers, social and family workers, the police and the military, the service sector, construction workers, etc.) and medical masks and gloves or equally safe protection should be provided to them by employers. Cloth masks should be used if medical masks are unavailable.
\item Countries should aim to eventually secure mass production and availability of appropriate medical masks (without exploratory valves) for the entire population during the pandemic.
\item Until supplies are sufficient, members of the general public should wear nonmedical fabric face masks when going out in public and medical masks should be reserved for essential functions.
\item The authorities should issue masking guidelines to residents and companies regarding the correct and optimal ways to make, wear and disinfect masks.
\item The introduction of mandatory masking will benefit from being rolled out together with campaigns, citizen initiatives, the media, NGOs, and influencers in order to avoid a public backlash in societies not culturally accustomed to masking. Public awareness is needed that ``masking protects your community—not just you''.\end{enumerate}
\section{Background}
\par Masks indisputably protect individuals against airborne transmission of respiratory diseases. A recent Cochrane meta-analysis found that masking, handwashing, and using gowns and/or gloves can reduce the spread of respiratory viruses, although evidence for any individual one of these measures is still of low certainty \citep{BurchBunt2020}. Currently, the lowest recorded daily growth rates in COVID-19 infections appear to be found in countries with a culture of mass face mask wearing, most of whom have also made mask wearing in public mandatory during the epidemic, and most of whom are not currently locked down—an observation that we study systematically in section~\ref{empircalsec}.
\par Outside of East Asia, support for universal masking is emerging elsewhere across the globe. The Czech Republic was the first non-Asian country to embrace and impose mandatory universal masking on March 11, 2020. The Czech policy swiftly inspired various initiatives from citizens, journalists and scientists—e.g., \citet{dekai2020}, \citet{howard2020}, \citet{manjoo2020_nytimes20200331}, \citet{abaluckchevalierchristakisformankaplankovermund2020}, \citet{FengShenXiaSongFanCowling2020}, \citet{fineberg2020}, \citet{tufekci2020_nytimes20200317}—and created global movements such as \#masks4all and \#wearafuckingmask. Their arguments build on the ability of the COVID-19 virus to spread from pre- and asymptomatic individuals who may not know that they are infected, and to linger in airborne droplets.
\par Leading political and medical experts who early were advocated masking included Chinese CDC director-general Prof. George Fu Gao \citep{servick2020}, former FDA commissioner Scott Gottlieb and Prof. Caitlin Rivers of Johns Hopkins \citep{gottliebrivers2020_washpost20200303}, and the American Enterprise Institute's roadmap \citep{gottliebriversmcclellansilviswatson2020}.
\par In early April 2020 a rapidly increasing number of governments from countries without a previous culture of mask wearing require or recommend universal masking including the Czech Republic, Austria and Slovakia. Additionally, public health bodies in the USA, Germany, France \citep{academienationaledemedecine2020}~and New Zealand have moved toward universal masking recommendations \citep{morgunovbaynomanawis2020}, as shown below in Figure \ref{maskingdata}.
\par The \citet{who2019} previously issued guidelines discouraging the use of masks in the public. However in early April 2020 the \citet{who2020_20200406} modified the guidelines, allowing self-made masks but rightly stressing the need to reserve medical masks for healthcare workers \citep{nebehayshalal2020_reuters20200403}, and to combine masking with the other main NPI needed to combat the pandemic. The policy shifts of the WHO and other CDCs reflect advances in our scientific understanding of this pandemic, and help legitimise the altruistic ``mask resistance'' of civil society in this global effort against COVID-19.
\section{SEIR modelling of universal masking impact}\label{seirsec}
\par In the first of our two new theoretical models, we employed stochastic dynamic network based compartmental SEIR modeling to forecast the relative impact of masking compared to the two main other societal non-pharmaceutical interventions, lockdown, and social distancing. 
\par The SEIR simulations were fit to the current timeline in many Western countries, with a lockdown imposed March the 24th (day 1) and planned to be lifted on May 31st. Universal masking is introduced in April. The simulation continues for 500 days from day 0, or around 17 months.
\par The experimental results strongly support the need for universal masking as an alternative to continued lockdown scenarios. For this strategy to be most effective, the vast majority of the population must adopt mask wearing immediately, as most regions outside East Asia are rapidly approaching Day 50.
\par In a SEIR model, the population is divided into compartments which represent different states with respect to disease progression of an individual: susceptible ($S$), exposed ($E$), infectious ($I$) and recovered ($R$). A susceptible individual may become exposed if they interact with an infectious individual at rate~ $\beta $~ (rate of transmission per $S$- $I$~ contact per time). From $E$~, the individual progresses to being infectious ($I$) and eventually recovered ($R$) with rates~ $\sigma $~ (rate of progression) and~ $\gamma $~ (rate of recovery), respectively. Additionally, individuals in $I$~are removed from the population (i.e., die of the disease) at rate~ ${\mu }_{I}$~ (rate of mortality). \newline  \newline We used a SEIR model implemented\footnote{\href{https://github.com/ryansmcgee/seirsplus}{https://github.com/ryansmcgee/seirsplus}} on a stochastic dynamical network that more closely mimics interactions between individuals in society, instead of assuming uniform mixing as is the case with deterministic SEIR models. Furthermore, such approach allows setting different model parameters for each individual, which we use to model masking. In a network model, a graph of society is built with nodes representing individuals and edges—their interactions. Each node has a state— $S$,~ $E$, $I$,~ $R$, or~ $F$ (the latter added to represent dead individuals). Adjacent nodes form close contact networks of an individual, while contacts made with an individual from anywhere in the network represent global contacts in the population. Varying the parameters affecting the two levels of interaction, as well as setting network properties such as the mean number of adjacent nodes (``close contacts'') allows us to model the degree of social distancing and lockdown measures. \newline  \newline Formally, each node $i$~is associated with a state ${X}_{i}$ which is updated based on the following probability transition rates: \newline  \newline \begin{equation}
\Pr({X}_{i}=S\to E)=[p\frac{\beta I}{N}+(1-p)\frac{\beta \sum _{j\in {C}_{G}(i)}{\delta }_{{X}_{j}=i}}{|{C}_{G}(i)|}]{\delta }_{{X}_{i}=S}\end{equation}\begin{equation}
\Pr({X}_{i}=E\to I)=\sigma {\delta }_{{X}_{i}=E}\end{equation}\begin{equation}
\Pr({X}_{i}=I\to R)=\gamma {\delta }_{{X}_{i}=I}\end{equation}\begin{equation}
\Pr({X}_{i}=I\to F)={\mu }_{I}{\delta }_{{X}_{i}=I}\end{equation}where~ ${\delta }_{{X}_{i}=A}=1$~ if the state of~ ${X}_{i}$~ is $A$~, or~ $0$~ if not, and where~ ${C}_{G}(i)$~ denotes the set of close contacts of node $i$~. 

\begin{figure*}
\centering
\includegraphics[width=0.9\textwidth]{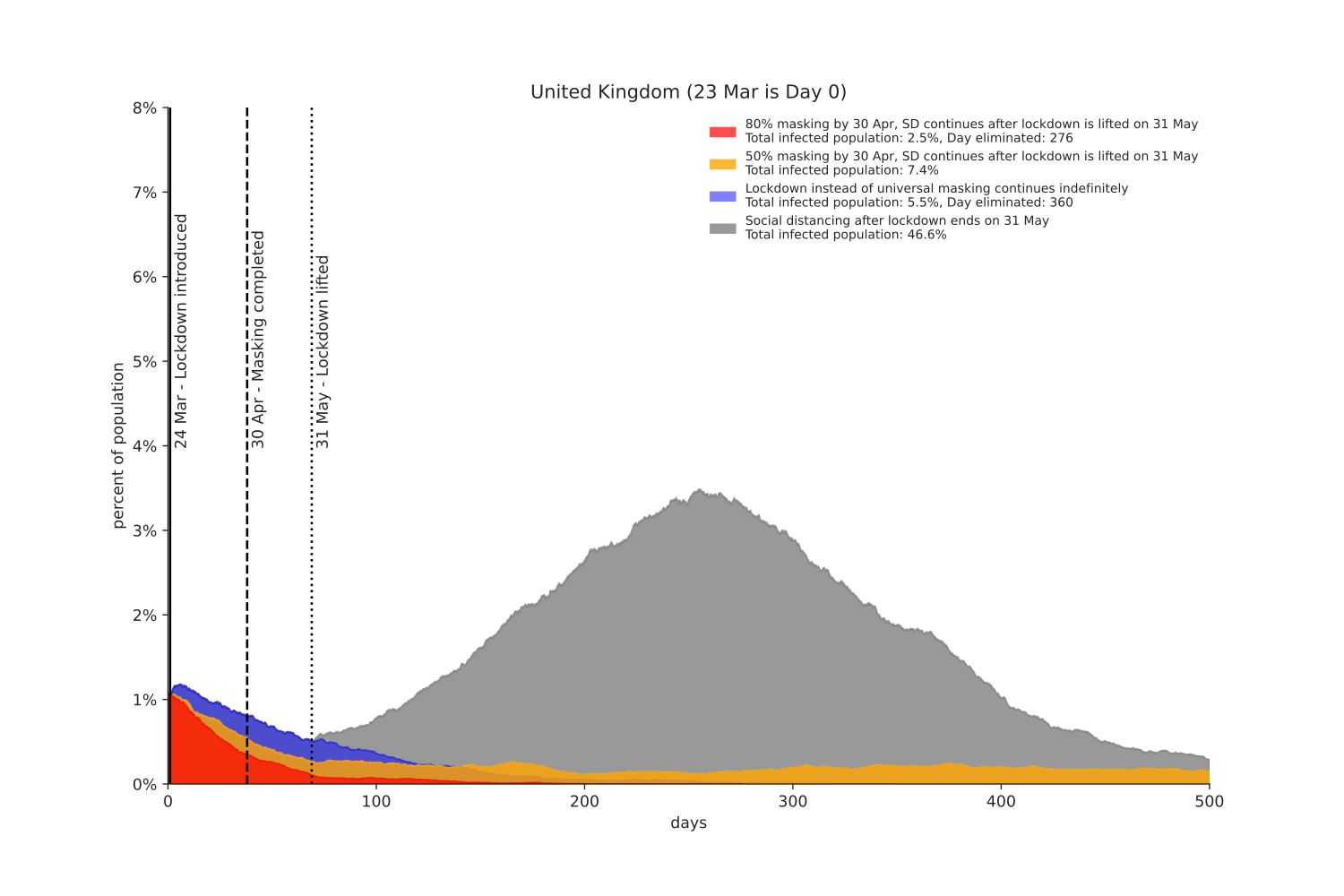}
\caption{Simulation results for a representative scenario: universal masking at 80\% adoption (red) flattens the curve significantly more than maintaining a strict lockdown (blue). Masking at only 50\% adoption (orange) is not sufficient to prevent continued spread. Replacing the strict lockdown with social distancing on May 31 without masking results in unchecked spread.}
\label{infections}
\end{figure*}
\begin{figure*}
\centering
\includegraphics[width=0.9\textwidth]{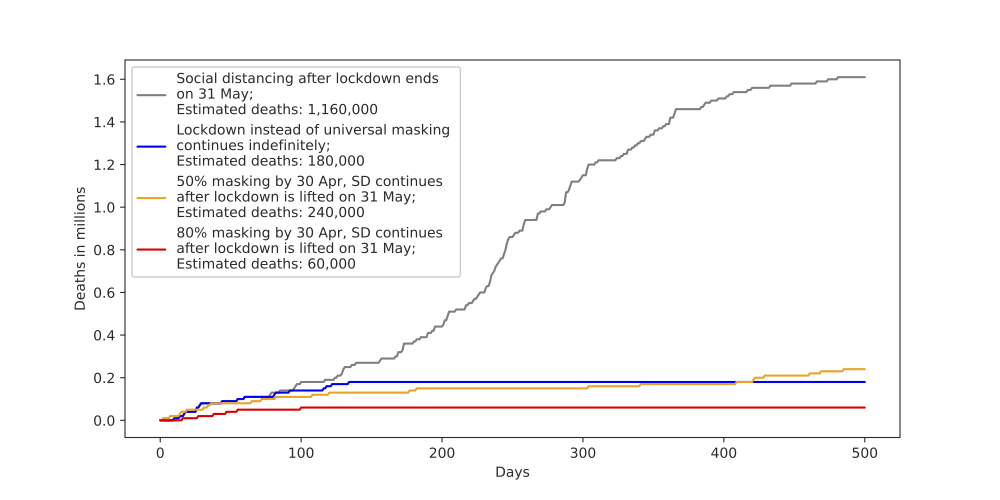}
\caption{Simulation results for a representative scenario: universal masking at 80\% adoption (red) results in 60,000 deaths, compared to maintaining a strict lockdown (blue) which results in 180,000 deaths. Masking at only a 50\% adoption rate (orange) is not sufficient to prevent continued spread and eventually results in 240,000 deaths. Replacing the strict lockdown with social distancing on May 31 without masking results in unchecked spread.}
\label{deaths}
\end{figure*}
\subsection{Experimental model}
\par We implemented SEIR dynamics on a stochastic dynamic network with a heterogeneous population. We assumed an initial infected population of 1\% and modelled the assumed effects of social distancing, lockdown, and universal masking over time on the rates of infection in the population.
\par All SEIR models were built using the SEIRS+ modelling tool\footnote{\href{https://github.com/ryansmcgee/seirsplus}{https://github.com/ryansmcgee/seirsplus}}, version 0.0.14. The baseline model parameters are fit to the empirical characteristics of COVID-19 spread, as documented in the SEIRS+ distributed COVID-19 notebooks. Specifically, we set $\beta =0.155$,~ $\sigma =1/5.2$ and~ $\gamma =1/12.39$. This parameterisation describes a SEIR model with best estimates for COVID-19 dynamics.
\par The initial infected population (${\mathrm{init}}_{i}$) was set to 1\%, and all others to 0\%. The size of the total population was set to 67,000 (a representative typical case, that is a factor of 1,000 from the population of the UK). 
\par \textbf{Social distancing.} In the model, social distancing was defined as the degree distribution of the contact network of an individual. Default interaction networks were used, constructed as Barabasi-Albert graphs with~$m=9$ and processes using the package function custom\_exponential\_graph with different scale parameters. Normal graph (scale=100) with mean degree 13.2, distancing graph (scale=10) with mean degree 4.1 and lockdown graph (scale=5) with mean degree 2.2.\textit{}
\par \textbf{Lockdown stringency.} Lockdown stringency was modelled considering no stringent lockdown (i.e. only social distancing) or stringent lockdown using the locality parameter p, which was set to 0.02 during lockdown and 0.2 during social distancing phases. This dictates the probability of individuals coming into contact with those outside of their immediate network. Assuming that individuals have around 13 contacts in normal everyday life, social distancing will reduce this to 4 and lockdown to only 2.
\par \textbf{Mask wearing.} A gradual increase in \textit{mask wearing }was modelled using a linear increase in the proportion of individuals randomly allocated with a reduced rate of transmission. The factor by which~$\beta $ was reduced was conservatively set to 2. The period of time over which the mask wearing went from 0 to maximum \% was set to 10 days. 50\% and 80\% maximum values were considered.
\par \textbf{Date fitting.} The progression in the number of deaths was used to fit the model to an approximate calendar date representing Day 0. For the representative typical case of the UK, this corresponded to Mar 23.
\subsection{Experimental results}
\par Figure~\ref{infections} shows the simulation results for a representative scenario: universal masking at 80\% adoption (red) flattens the curve significantly more than maintaining a strict lockdown (blue). Masking at only 50\% adoption (orange) is not sufficient to prevent continued spread. Replacing the strict lockdown with social distancing on May 31 without masking results in unchecked spread.
\par Our model suggests a substantial impact of universal masking. Without masking, but even with continued social distancing in place once the lockdown is lifted, the infection rate will increase and almost half of the population will become affected. This scenario, rendered in grey in Figure~\ref{infections}, would potentially lead to over a million deaths in a population the size of the UK. A continued lockdown, illustrated in blue colour, does eventually result in bringing the disease under control after around 6 months. However, the economic and social costs of a ``full body lockdown'' will be enormous, which strongly supports finding an alternative solution.
\par In the model, social distancing and masking at both 50\% and 80\% of the population—but no lockdown beyond the end of May—result in substantial reduction of infection, with 80\% masking eventually eliminating the disease. Figure~\ref{deaths} shows the simulation results for a representative scenario: universal masking at 80\% adoption (red) results in 60,000 deaths, compared to maintaining a strict lockdown (blue) which results in 180,000 deaths. Masking at only a 50\% adoption rate (orange) is not sufficient to prevent continued spread and eventually results in 240,000 deaths. Replacing the strict lockdown with social distancing on May 31 without masking results in unchecked spread.
\section{Agent based modelling of universal masking impact}\label{abmsec}
\par In the second of our two new theoretical models, we employed stochastic individual agent based modelling (ABM) as an alternative~Monte Carlo simulation technique for understanding the impact of universal masking. Agent based models have roots in various disciplines. A stochastic agent program can be defined as a agent function~ $f:\mathbf{p}\to \Pr(a)$~ which maps possible \emph{percept} vectors to a probabilistic distribution over possible \textit{actions} (or to states that influence subsequent actions). In AI, \citet{russellnorvig2009} summarise five classes of intelligent agents: simple reflex agents, model-based reflex agents, goal-based agents, utility-based agents, and learning agents; note, however, that agents may also be susceptible to \emph{imperceptible} environmental factors such as viruses. \citet{hollandmiller1991} discuss artificial adaptive agents for modeling complex systems in economics. \citet{bonabeau2002} surveys agent based models for simulating human systems. 
\par As in other disciplines, ABM approaches in epidemiology (see, e.g., \citet{huntermacnameekelleher2017}. \citet{tracycerdakeyes2018}, or \citet{huntermacnameekelleher2018}) have several advantages compared to compartmental models which group undifferentiated individuals into large aggregates (like in the above SEIR simulation). First, because the behavior and characteristics of each agent is independent, they can simulate complex dynamic systems with less oversimplification of rich variation among individuals. Second, because agents can be simulated in physical two- or three-dimensional spaces, they can better simulate the geometry of contact between individuals, which is highly relevant in epidemiology. Third, the randomization on each run makes the statistical \emph{variance} more apparent than in the SIR family of models, whose smooth curves often misleadingly convey more certainty than warranted. Fourth, ABMs lend themselves well to visualization, as seen in Figure~\ref{abmwebapp}, which helps convey the non-linear behavior of complex dynamic systems—an especially relevant advantage when the exponential effect of masking can be counterintuitive in many cultures due to pre-existing cultural biases \citep{Leung2020_Time20200312} and unconscious cognitive biases \citep{dekai2020}.
\subsection{Mask characteristics}
\par The ABM approach allows us to put masks on individual agents and to assign properties to those masks, to shed light on the question of how face masks—even nonmedical cloth masks—carry the promise to be so surprisingly effective. The objective is to examine how even a small barrier to individual infection transmission can multiply into a substantial effect on the level of communities and populations.
\par Face masks work in two ways: They can protect an infected person from spreading the virus (transmission), and they can limit how much the non-infected individual is exposed to the virus (absorption). Traditionally, masks are worn to protect the wearer from being infected by an ill person when in close and prolonged contact. In such classic situations, for instance in hospitals and elderly homes, only medical masks combined with other protective equipment provide protection. Comparing different mask materials, medical masks have been found to be up to three times more effective in blocking transmission compared to homemade masks~\citep{DaviesThompsonGiriKafatosWalkerBennett2013}.~Surgical masks most efficaciously reduce the emission of influenza virus particles into the environment in respiratory droplets. Still, although masks vary greatly in their ability to protect, using any type of face mask (without an exploratory valve) can help decrease viral transmission \citep{SandeTeunisSabel2008}.
\par However, the effect of universal masking does not require full protection from disease to be effective in lowering infection rates of COVID-19.~Masks may be especially crucial for containing the COVID-19 pandemic, since many infections appear to come from people with no signs of illness. For instance, around 48\% of COVID-19 transmissions were pre-symptomatic in Singapore and 62\% in Tianjin, China~\citep{ganyanikremerchentornerifaeswallingahens2020}. This suggests that masking needs to be universal and not restricted to individuals who think they may be infected. 
\begin{figure*}
\centering
\includegraphics[width=1\textwidth]{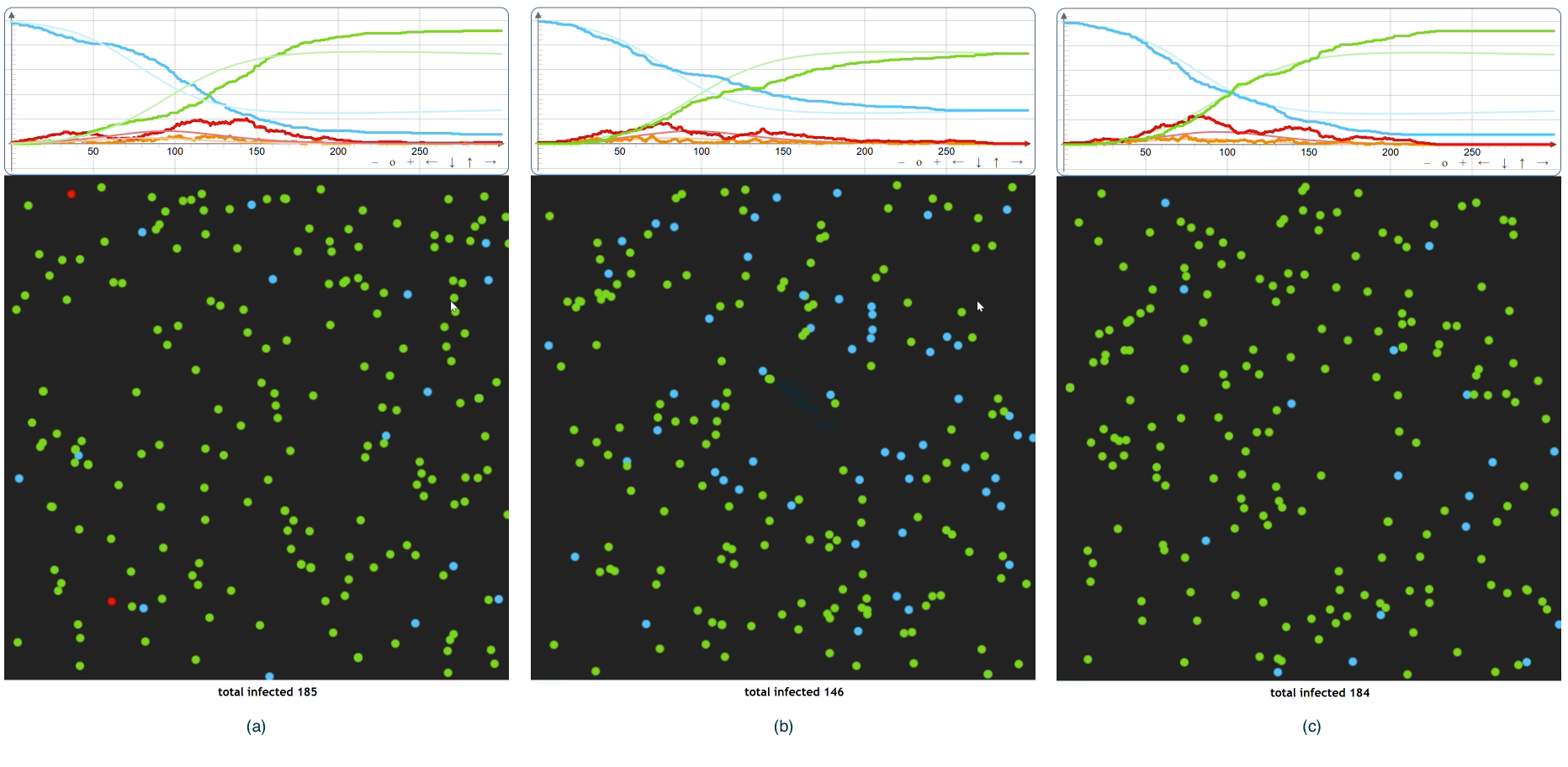}
\caption{Three successive randomised runs of the agent based model for 300 days, with no mask wearing. Blue is susceptible, orange is exposed, red is infected, and green is recovered. The contrastive SEIR baseline model's predicted curves are shown in thinner, fainter lines. The ABM runs produce curves with a fine granularity of randomisation, centering on average around the ODE based SEIR curves.}
\label{abmbaseline}
\end{figure*}
\begin{figure*}
\centering
\includegraphics[width=0.65\textwidth]{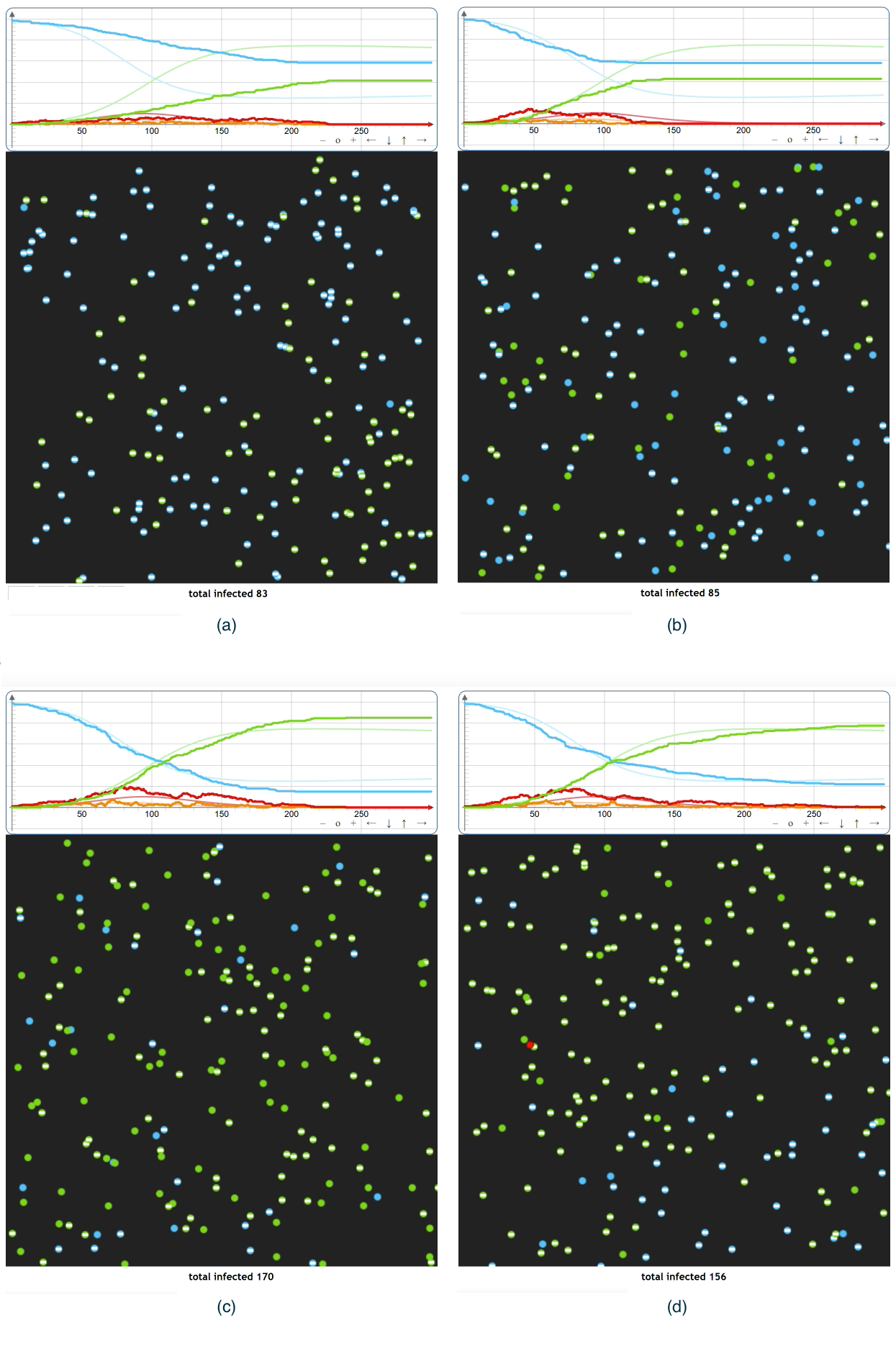}
\caption{Four ABM runs under varying masking scenarios. (a) 100\% of the population wearing masks from the onset of the outbreak, with excellent suppression of infection spread. (b) 0\% of the population initially wearing masks, but instituting near universal masking of 90\% of the population at day 50, still with significant suppression of infection spread. (c) 0\% of the population initially wearing masks, and instituting some masking of 50\% of the population at day 50, with not much impact on infection spread. (d) 0\% of the population initially wearing masks, but instituting near universal masking of 90\% of the population at day 75 with not much impact on infection spread.}
\label{abmscenarios}
\end{figure*}
\par Furthermore, the SARS-CoV-2 virus is known to spread through airborne particles \citep{LeungChuShiuChanMcDevittHauYenLiIpPeirisSetoLeungMiltonCowling2020} and quite possibly via aerosolised droplets as well according to \citet{Service2020}, \citet{VanDoremalenBushmakerMorrisHolbrookGambleWilliamsonTaminHarcourtThornburgGerberLloyd2020}, \citet{SantarpiaRiveraHerreraMorwitzerCreagerSantarpiaCrownBrettmajorSchnaubeltBroadhurstLawlerReidLowe2020}, and \citet{LiuNingChenGuoLiuGaliSunDuanCaiWesterdahlLiuHoKanFuLan2020}. It may linger in the air for and travel several meters, which is why social distancing rules require at least 2 meters between individuals to be effective.
\subsection{Experimental model}
\par As a contrastive baseline we employed a compartmental SEIR model with the same parameters as given~for our SEIR experiments of section~\ref{seirsec}.
\par For the new agent based model, we implemented an environment consisting of a square wraparound two-dimensional space, within which a population of individual agents reside in four states: susceptible ($S$), exposed ($E$), infectious ($I$) and recovered ($R$). The wraparound space means that agents who move outside a border re-enter the square from the opposite side. As in our SEIR models, the initial infected population (${\mathrm{init}}_{i}$) was set to 1\%, and all others to 0\%. The size of the total population was set to 200, but the wraparound feature of the two-dimensional space in effect represents arbitrarily larger spaces that are approximated by replicated square tiles, thus giving more accurate dynamics without boundary effects from small spaces. 
\par To best fit the same empirical characteristics of COVID-19 spread as our SEIR models, we again set~ $\sigma =1/5.2$ and~ $\gamma =1/12.39$. Note that $\beta $ is inapplicable in the ABM since infection transmission between individuals arises from physical proximity, which is more realistic than randomly infecting other individuals anywhere with some probability $\beta $~with no regard to their physical location. In the baseline Monte Carlo simulation, agents decide on a random destination location within a parameterised radius of their current point, then proceed at a parameterised speed to move there, and then repeat the process iteratively. We adjusted such ABM-specific parameters, as well as physical exposure distance, to optimise fit to the baseline SEIR model curves, assuming none of the population to be wearing masks. Again, this was done so as to best approximate known COVID-19 dynamics. 
\par ABM runs were for 300 days from the onset of the outbreak since empirically, the emergent SEIR curves stabilise before the 300th day.
\par To model the impact of masking, the following masking parameters can be varied:
\par \textbf{Mask wearing.} Gradual increases (or decreases) in \textit{mask wearing }can be modelled using parameterised rates of masking $M $(or unmasking $U$) in the proportion of unmasked (or masked) individuals. The parameters ${m}_{min}$ and ${m}_{max}$ also allow modelling the minimum and maximum absolute numbers of masked agents. These masking parameters can be dynamically adjusted any time during any ABM run, to simulate varying policy decisions and cultural mindset shifts. 
\par \textbf{Mask characteristics.} Varying degrees of mask effectiveness are modelled by the \textit{mask transmission rate}~$T$ and \textit{mask absorption rate}~$A$, which denote the proportion of viruses that are stopped by the mask during exhaling (transmission) versus inhaling (absorption), respectively. We set $T=0.7$ and $A=0.7$ to model the use of inexpensive, widely available, and even nonmedical or homemade masks with only 70\% effectiveness for universal masking, and not higher quality N95, N99, N100, FFP1, FFP2, or FFP3 masks which in many regions need to be reserved for medical staff. 
\begin{figure*}
\centering
\includegraphics[width=0.65\textwidth]{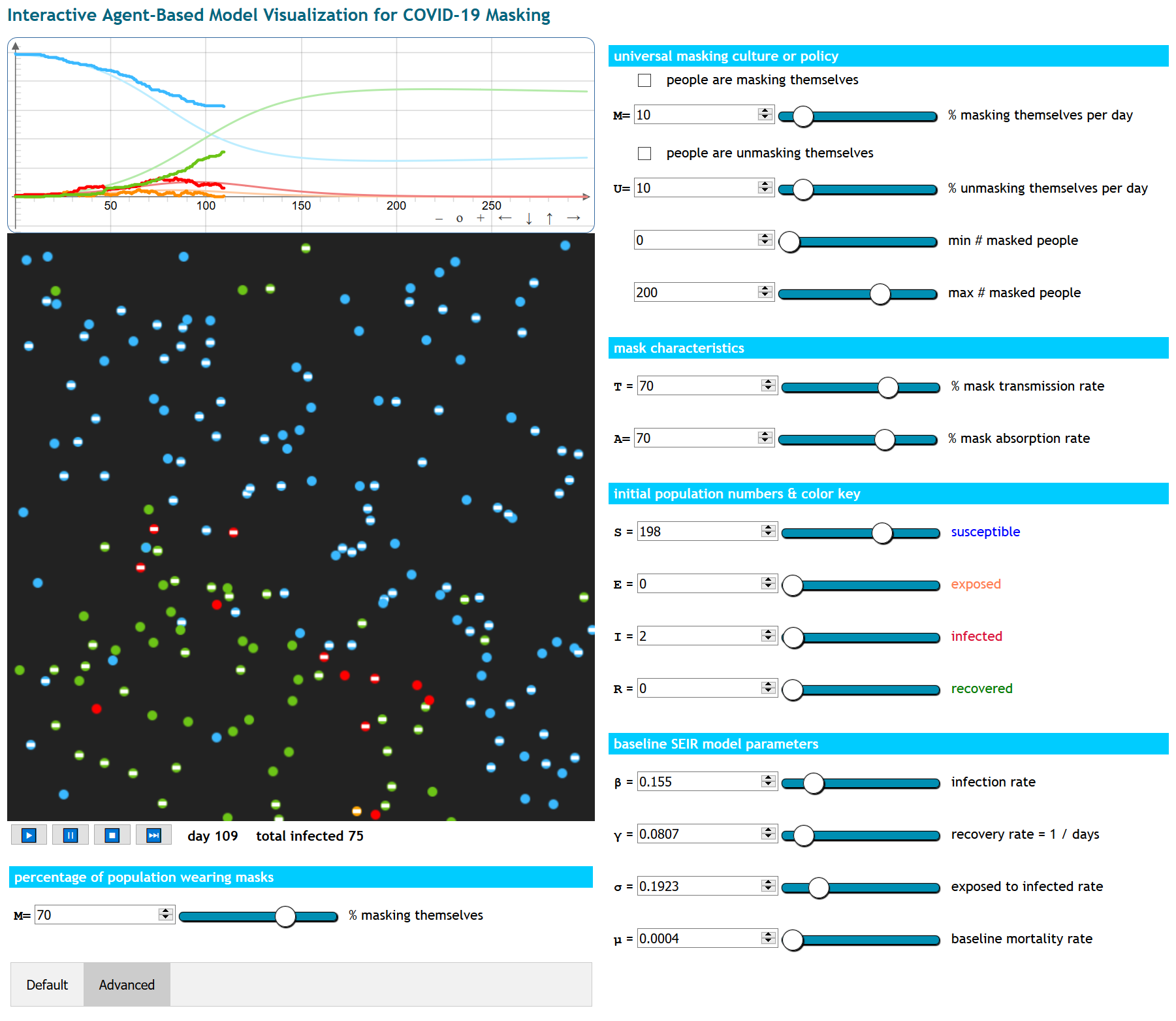}
\caption{Interactive visualisation tool for the ABM simulation model to help policy makers and the general public gain a more concrete feel for how masks impact the dynamics of virus spread, available online at http://dek.ai/masks4all.}
\label{abmwebapp}
\end{figure*}
\subsection{Experimental results}
\par ABM simulation shows that universal masking can significantly reduce virus spread if adopted sufficiently early, even if the masks are nonmedical or homemade.
\par Figure~\ref{abmbaseline} shows three successive runs for the baseline $m=0$~case with zero mask adoption. Each dot (which is in motion during simulation runs) represents an individual agent, who may become exposed to the virus through proximity to other agents who are infectious. Blue dots are healthy susceptible agents, orange dots are exposed agents, red dots are infected agents, and green dots are recovered agents. A dot with a white rectangle on it represents an agent who is wearing a mask. 
\par The three baseline ABM runs show how chance plays a significant role in the dynamics of virus spread. Since each simulation run is randomised, to decrease variance requires observation over multiple runs. On average, the baseline case with zero mask adoption adheres to the simpler SEIR model's predicted curves.
\par Figure~\ref{abmscenarios} compares typical runs for four scenarios that simulate how COVID-19 spreads among individual agents under different masking scenarios, with the contrastive baseline SEIR model curves shown in thin lines as a reference: (a) ${m}_{0}=\mathrm{100\%}$ meaning that the entire population adopts mask at the onset of the outbreak on day 0; (b) ${m}_{0}=\mathrm{0\%},{m}_{50}=\mathrm{90\%}$ meaning that none of the population is wearing masks at the onset but that nearly universal masking is instituted on day 50; and (c) ${m}_{0}=\mathrm{0\%,}{m}_{50}=\mathrm{50\%}$ meaning that none of the population is wearing masks at the onset but that half of the population adopts masks on day 50, and (d) ${m}_{0}=\mathrm{0\%},{m}_{75}=\mathrm{90\%}$ meaning that none of the population is wearing masks at the onset but that nearly universal masking is instituted on day 75. 
\par In scenario (a), a dramatic decrease in the number of infections is evident as a result of universal masking at the onset of the outbreak. Unfortunately, most regions outside East Asia missed the time window for scenario (a).
\par In scenario (b), even though the population is not initially wearing masks, if universal masking is instituted by day 50, good chances of dramatic suppression of infection rates can still be achieved. Fortunately, this option is within reach of most regions at the time of writing. 
\par In scenario (c), again the population is not initially wearing masks. On day 50, half the population dons masks, but unlike scenario (b) which succeeds with 90\% universal masking, unfortunately 50\% is an insufficient level of mask adoption to suppress infection rates to a significant degree.
\par In scenario (d), the population again is not initially wearing masks, but unlike scenario (b) the 90\% universal masking is not instituted until day 75, instead of day 50. Waiting too long unfortunately greatly decreases the degree to which infection rates can be suppressed.
\par To help policy makers and the general public gain a more concrete feel for how masks impact the dynamics of virus spread, we have made available online\footnote{\href{http://dek.ai/masks4all}{http://dek.ai/masks4all}} an interactive visualisation tool for the ABM simulation model, as shown in Figure~\ref{abmwebapp}. The default view allows direct adjustment in real time of the percentage of masked individual agents through a slider control. Optional advanced controls allow playing with various scenarios: whether masking is used, the adoption rate of masking, virus transmission and absorption rates through masks of varying quality, as well as other modelling parameters such as the initial numbers of susceptible, exposed, infected, or recovered agents, and the contrastive baseline SEIR model parameters. 
\section{Evaluation of model predictions against empirical data on universal masking impact}\label{empircalsec}
\par For validation of the foregoing SEIR and ABM predictive models it is necessary to compare against what little historical macro scale empirical data is available, since precise numbers~are not yet known for masking rates, mask transmission and absorption rates, and infectious but asymptomatic cases.
\subsection{Validation data set}
\par We collected a new data set describing the degree of success in managing COVID-19 by~countries or regions segmented by the prevalence or enforcement of universal masking. The data set covers (a)~a selection of 38 countries or provinces in Asia, Europe and North America that have similar, high levels of economic development (based on World Bank GDP purchasing power parity per capita), (b)~detected COVID-19 cases from Jan 23 to April 10, 2020, and (c)~characteristics of universal masking culture and/or universal masking orders or recommendations by governments.
\begin{figure}
\centering
\includegraphics[width=0.8\columnwidth]{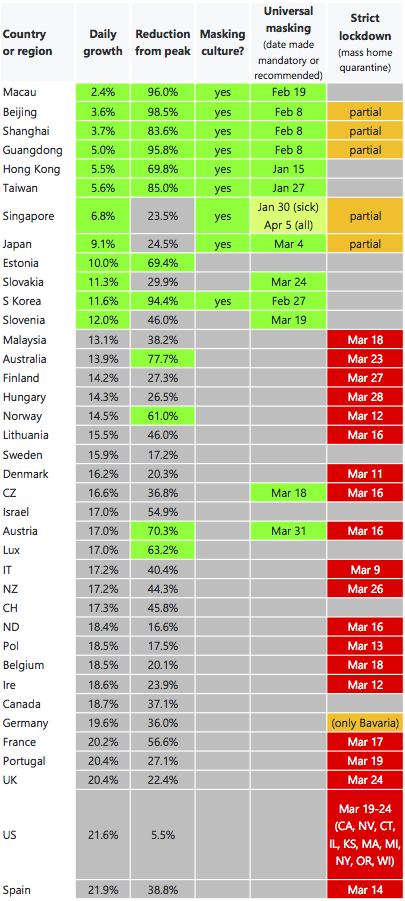}
\caption{Epidemic daily growth and reduction from peak daily growth, together with masking culture, universal masking policy, and lockdown policy, from January 23 to April 10, 2020 for selected list of countries or provinces with high GDP PPP per capita in Asia, Europe and North America. Universal masking was employed in every region that handled COVID-19 well. Sources: John Hopkins, Wikipedia, VOA News, Quartz, Straits Times, South China Morning Post, ABCNews, Time.com, Channel New Asia, Moh.gov.sg, Reuters, Financial Times, Yna.co.kr, Nippon.com, Euronews, Spectator.sme.sk}
\label{maskingdata}
\end{figure}
\begin{figure*}
\centering
\includegraphics[width=0.9\textwidth]{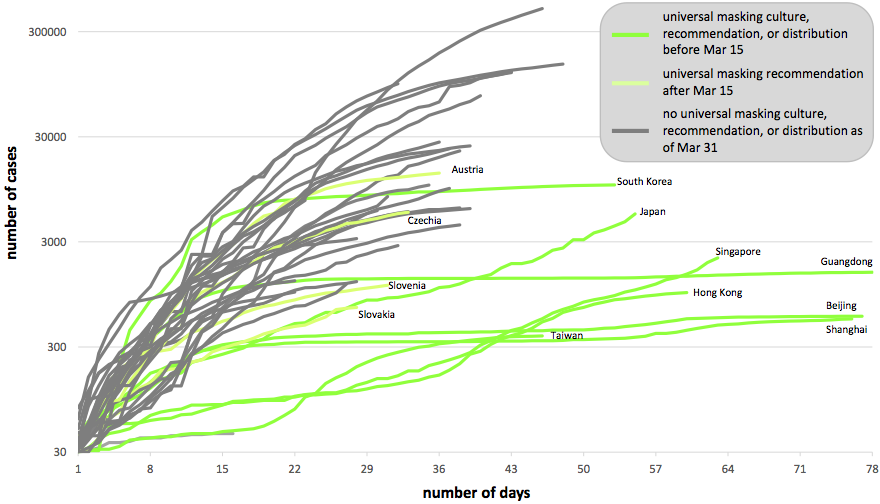}
\caption{Daily growth curves showing the impact of universal masking on epidemic control: epidemic trajectory after 30 detected cases in universal masking selected countries and provinces (green) vs. others (grey). Masking is nearly perfectly correlated with lower daily growth or strong reduction from peak growth of COVID-19. Sources: John Hopkins, Wikipedia, VOA News, Quartz, Straits Times, South China Morning Post, ABCNews, Time.com, Channel New Asia, Moh.gov.sg, Reuters, Financial Times, Yna.co.kr, Nippon.com, Euronews, Spectator.sme.sk}
\label{maskingcurves}
\end{figure*}
\subsection{Feature extraction}
\par From our data set's 38 selected countries, we computed (a) the daily growth of confirmed cases, as well as (b) reduction from peak of new cases. Sorted in increasing order of the daily growth, Figure \ref{maskingdata} presents these figures alongside features extracted from our data set denoting each country or region's (c) masking culture, (d) universal masking policy, and (c) lockdown policy. Additional clarification on definitions of a couple of these features follow.
\par \textbf{Masking culture} is defined as an established practice by a significant section of the general population to wear face masks prior to the start of the Covid-19 pandemic. A cursory review of the scientific literature and the general press has identified Japan, Thailand, Vietnam \citep{burgesshorii2012}, China's urban centers \citep{kuo2014},~Hong Kong \citep{cowlingalingtsanglifongliaokwanleechiuwuwuleung2020}, Taiwan, Singapore and South Korea (\citet{yang2014}, \citet{jennings2020}) as countries with such a consistent practice, at least in the decade predating the Covid-19 pandemic. Nevertheless, the notion of "culture" should not imply that the practice of face mask wearing has been extensive and consistent throughout time. For example, though this practice may have fit with preexisting Taoist and health precepts of Chinese traditional medicine, its actual emergence may be relatively recent, starting with the industrialization of Japan at the start of the XXth century and both the flu pandemics of the XXth century as well as the rise of particle pollution \citep{yang2014}. The rest of the above-listed east Asian countries has followed the same course in the second half of the XXth century, including China as it was confronting a severe particle pollution crisis in the first part of the 2010s (\citet{kuo2014}, \citet{li2014}, \citet{hanssteinechegaray2018}). Beyond price, availability and government recommendation, the actual practice of masking in the Asian general population may be mediated by factors such as social norms or peer-pressure, perception of one's competence, past behaviors or perception of the danger \citep{hanssteinechegaray2018}. As an example of the latter, in Hong Kong, masking was practiced by 79\% of the general population during the 2003 SARS outbreak, but by only a maximum of 10\% of the general population during the Influenza A pandemic in 2009 \citep{cowlingalingtsanglifongliaokwanleechiuwuwuleung2020}.
\par \textbf{Universal masking policy.} Additionally, to the extent that government recommendations or mandatory orders may shape perceptions and assist in masks availability, it may amplify the masking practice in the general population. It can thus be assumed that the maximum potency of universal masking in the context of epidemics may be reached when a government issues a mandatory or highly recommended order to the general population, issued at an early date, supported by the availability of face masks and amplified by a pre-existing "masking culture". In that case, we make the reasonable assumption that such national situations may be used to validate our SEIR and ABM predictive models at maximum values (80-90\%) for the percentage of the general population wearing masks.
\par We also computed two additional meta-features to classify \emph{successful} management of the epidemic outbreak. These meta-features help to highlight both (a)~success in suppressing growth from the start (e.g., Hong Kong or Taiwan) or (b)~success in managing the epidemic by reducing the number of new cases after a peak (e.g., South Korea).
\par \textbf{Successful suppression of daily growth} is defined as being~below 12.5\% daily growth (equivalent to number of cases doubling at the slower pace of 6 days or more) once the number of detected cases first reached 30. These daily growth rates are highlighted in green in Figure \ref{maskingdata}.
\par \textbf{Successful reduction from peak} is defined as a recent, significant ($>$60\%) reduction of new cases calculated as the average of the last five days before April 10, 2020 compared to the average of the three highest number of daily new cases up to~April 10, 2020 starting from the date when the number of detected cases first reached 30. Again, these reductions from peak are highlighted in green in Figure \ref{maskingdata}.
\subsection{Validation results}
\par Results bear out the predictions made by our SEIR and agent-based models as described in sections~\ref{seirsec} and~\ref{abmsec}.
\par In Figure \ref{maskingdata}, the green (successful supression of daily growth and/or reduction from peak) areas show that as of April 10, 2020, an overwhelming majority of countries or regions that have best managed COVID-19 outbreaks were countries or regions with either (1)~established universal masking cultures or (2)~mandatory orders or government recommendations supported by significant and early mask production destined for the general population. These countries or regions include Taiwan, South Korea, Singapore, Japan, autonomous special administrative regions such as Hong Kong or Macau, and Chinese provinces such as Beijing, Shanghai, or Guangdong. In effect, masking in public has been required in Taiwan, metropolitan areas in China such as Shanghai and Beijing (as well as Guangzhou, Shenzhen, Tianjin, Hangzhou, and Chengdu), Japan, South Korea, and other countries \citep{morgunovbaynomanawis2020}. On the other hand, the red (strict lockdown without universal masking) areas show that most of the countries which have adopted mass testing, tracking and quarantining, but lack a universal masking culture and clear recommendations and availability for universal masking, have not achieved an equivalent level of COVID-19 epidemic control as of April 10, 2020. This nearly perfect correlation between early universal masking and successful management of COVID-19 outbreaks bears out our SEIR and ABM predictions.
\par In Figure \ref{maskingcurves}, daily growth curves were extracted from our data set in order to reveal the impact of universal masking on epidemic control on a time axis. Results show that universal masking is nearly perfectly correlated with lower daily growth rates of COVID-19 cases over time, again validating the predictions from our SEIR and agent based models.
\par
\begin{figure*}
\centering
\includegraphics[width=0.9\textwidth]{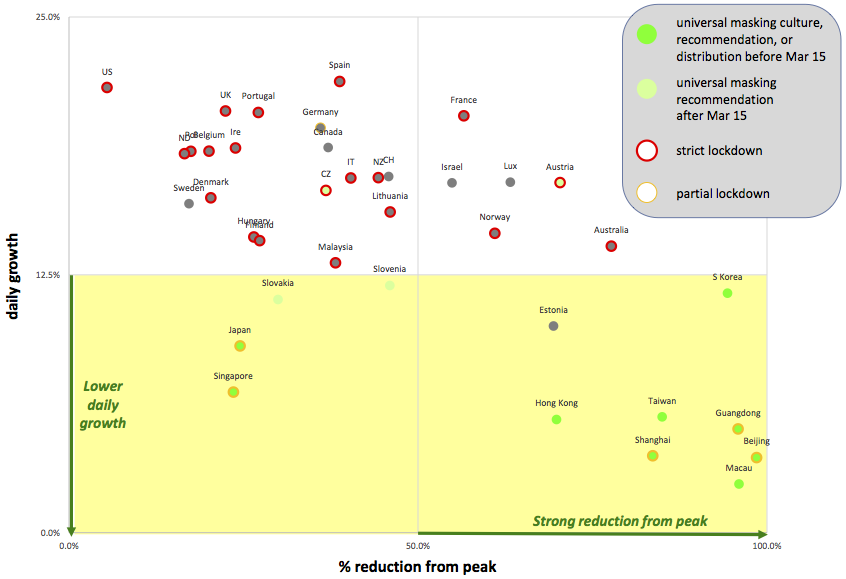}
\caption{Visual representation of epidemic daily growth versus percentage reduction from peak daily daily growth in quadrants showing the impact of universal masking on epidemic control: and reduction from peak, from January 23 to April 10, 2020 for selected list of countries or provinces with high GDP PPP per capita in Asia, Europe and North America. Masking is nearly perfectly correlated with lower daily growth or strong reduction from peak growth of COVID-19. Sources: John Hopkins, Wikipedia, VOA News, Quartz, Straits Times, South China Morning Post, ABCNews, Time.com, Channel New Asia, Moh.gov.sg, Reuters, Financial Times, Yna.co.kr, Nippon.com, Euronews, Spectator.sme.sk}
\label{maskingquadrants}
\end{figure*}

 In Figure \ref{maskingquadrants}, daily growth was plotted against versus percentage reduction from peak daily daily growth. Green points, representing countries or regions with early universal masking, disproportionately fall within the two lower quadrants which represent successful management of COVID-19 outbreaks. Red points, representing countries with strict lockdowns but not universal masking, nearly all fall in the two upper quadrants which represent less successful management of COVID-19 outbreaks. Light green points, representing countries or regions with late universal masking, tend to fall in the middle regions. Again, the strong correlation of universal masking with successful control of COVID-19 case growth bears out our SEIR and agent based models' predictions. 
\par \textbf{Validation of the need for \emph{universal} masking.} These validations highlight the gradual nature of the protection against COVID-19 achieved with a higher fraction of the population practicing masking, as observed in the SEIR and ABM simulations when comparing situations with 80-90\% universal masking versus only 50\% masking or none. In countries or provinces with masking culture and universal masking orders or recommendations before March 15, 2020, the average daily growth was 5.9\% and the reduction from peak was 74.6\%. In the countries without masking culture and~universal masking orders or recommendations after March 15, 2020, the average daily growth was 14.2\% and the reduction from peak was 45.8\%. Finally, for the rest of the other countries,~the average daily growth was 17.2\% and the reduction from peak was 37.4\%, the lowest results of the sample.~The latter group includes countries that have gone into "strict lockdown" (or mass home quarantine) for 20 out of 27 countries (74\%). This is much higher than for the intermediate group of countries without masking culture and "late" universal masking orders (2 out 4, or 50\% of the sample), or the first group of countries and provinces with masking culture and "early" universal masking orders. In that first group, no countries or provinces had to endure "strict lockdown". 
\par \textbf{Validation of the need for \emph{early} universal masking.} Yet even within this first group, the strength of \emph{early} universal masking recommendations from the government may impact the proportion of the general population actually wearing masks and thus the level of epidemic control, as per our models' SEIR and ABM predictions. For example, Singapore initially encouraged people to wear masks only when feeling unwell. Then, on April, 5, the government changed policy and decided to distribute reusable face masks to all households \citep{cheong2020}. On the other end, Hong Kong decided by January 24, 2020 to advise the general population to wear surgical masks in crowded places and public transports \citep{hkdoh2020}. As can be observed from~Figure~\ref{maskingdata}, as of April 10, 2020, the characteristics for epidemic control in terms of daily growth and peak from reduction are better for Hong Kong than for Singapore. These variations may be related to levels of adherence to masking by the general population. Though there are no available data as of April 10, 2020 as per adherence to universal masking in Singapore, telephone surveys in Hong Kong done in February 11-14, 2020 and then in March 10-13, 2020, both after Department of Health public advice, have shown declared masking adherence at the very high levels of 97.5\% and 98.8\% respectively when going out \citep{cowlingalingtsanglifongliaokwanleechiuwuwuleung2020}. Assuming the adherence level to masking was lower in Singapore since the general population order came much later, this would support our SEIR and ABM predictions of the need for \emph{early} institution of universal masking.
\par Although these correlations may also be sensitive to other unobserved factors, the theoretical SEIR and ABM predictions as empirically validated in the various ways described here call for urgent policy and public action even as further enquiry is pursued into the effects of masking. Our results also confirm and amplify other previous findings. A recent macro-level regression analysis by economists at Yale University, taking into account masking cultures and times of country COVID-19 policy responses, estimated that growth of COVID-19 rates only half that of mask wearing countries—the growth rate of confirmed cases is 18\% in countries with no pre-existing mask norms and 10\% in countries with such norms, while the growth rate of deaths is 21\% in countries with no mask norms and 11\% in countries with such norms. The authors note that such a 10\% reduction in transmission probabilities could correspond to a per capita gain of \$3,000-6,000 per each additional cloth mask, and that the economic benefits of each medical mask for healthcare personnel could be substantially larger \citep{abaluckchevalierchristakisformankaplankovermund2020}.
\section{Conclusion: Universal masking needs broad support and clear guidelines}
\par Our SEIR and ABM models suggests a substantial impact of timely universal masking. Without masking, but even with continued social distancing in place once the lockdown is lifted, the infection rate will increase and almost half of the population will become affected. This scenario would potentially lead to over a million deaths in a population the size of the UK. Social distancing and masking at both 50\% and 80-90\% of the population—but no lockdown beyond the end of May—result in substantial reduction of infection, with 80-90\% masking eventually eliminating the disease.
\par Moreover, for a significant chance of mitigating infection growth rates, universal masking must be adopted early—by day 50 from the onset of COVID-19 outbreaks.
\par Without masking, lifting lockdown after nine weeks while keeping social distancing measures will risk a major second wave of the epidemic in 4-5 months' time. However, if four out of five citizens start wearing cloth masks in public before the lockdown is lifted, the number of new COVID-19 cases could decline enough to exit lockdown and still avoid a second wave of the epidemic. If only every second person starts wearing a mask, infection rates would also decline substantially, but likely not by enough to prevent the second wave.
\par Combined with the correlational empirical evidence, our results highlight the need for mass masking as an alternative to a continued lockdown scenario. For this strategy to be most effective, the vast majority of the population needs to adopt mask wearing immediately. When a well-timed ``mouth-and-nose lockdown'' accompanies the current ``full body lockdown'', both the human and economic costs of the COVID-19 pandemic can be significantly lowered. 
\par Our theoretical and empirical results are in line with previous studies suggesting that a high rate of masking may be needed in a population to provide efficient protection from influenza \citep{YanGuhaHariharanMyers2019}~and that masking can be an effective intervention strategy in reducing the spread of a pandemic \citep{TrachtValleHyman2010}.
\par Furthermore, universal masking can reduce stigmatization of ethnic groups, risk groups, or the sick and contribute to public solidarity \citep{FengShenXiaSongFanCowling2020}.
\par We urge governments and international bodies who have not yet done so to consider masking as one of the key tools in population policy after the COVID-19 lockdowns and until the virus is under control. The analysis presented here supports recent studies \citep{abaluckchevalierchristakisformankaplankovermund2020}, suggesting that the effectiveness of universal masking is comparable to that of social distancing or a societal lockdown with closed workplaces, schools, and public spaces and limited geographical mobility. The results from our simulation help explain the dynamics behind the perplexing advantage in the Asian experience of tackling COVID-19 compared to the situation elsewhere.
\par Our analyses lead to the following key policy recommendations:
\begin{enumerate}
\item Masking should be mandatory or strongly recommended for the general public when in public transport and public spaces, for the duration of the pandemic.
\item Masking should be mandatory for individuals in essential functions (health care workers, social and family workers, the police and the military, the service sector, construction workers, etc) and medical masks and gloves or equally safe protection should be provided to them by employers. Cloth masks should be used if medical masks are unavailable.
\item Countries should aim to eventually secure mass production and availability of appropriate medical masks (without exploratory valves) for the entire population during the pandemic.
\item Until supplies are sufficient, members of the general public should wear nonmedical fabric face masks when going out in public and medical masks should be reserved for essential functions.
\item The authorities should issue masking guidelines to residents and companies regarding the correct and optimal ways to make, wear and disinfect masks.
\item The introduction of mandatory masking will benefit from being rolled out together with campaigns, citizen initiatives, the media, NGOs, and influencers in order to avoid a public backlash in societies not culturally accustomed to masking. Public awareness is needed that ``masking protects your community—not just you''.\end{enumerate}
\par The effectiveness of universal masking in a given population is likely to depend on (a) the type of masks used, (b) the acceptance of masking in the population, (c) the level of contagion of the virus, and (d) what other interventions have been applied. From this perspective, the Central European experience will be highly informative, since it represents the first major shift to universal masking in a formerly non-masking culture. The effects of this pioneering intervention on infection rates and fatalities will appear only in the forthcoming weeks, although Slovakia and Slovenia are currently showing early indications of progress (see Figure \ref{maskingcurves}). In any case, they illustrate that a country with no prior history of mask wearing in public may rapidly change course, and quickly adopt masks as a non-stigmatised—even street smart—way to express caring and solidarity in the community.
\par The medical and social risks of increased infections need to be countered by proper advice in the public domain. Some studies do indicate negative effects of naive improper cloth mass use, for instance higher risks of infection due to moisture retention, reuse of poorly washed cloth masks, and poor filtration in comparison to medical masks \citep{macintyresealedunghienngachughtairahmandwyerwang2015}. To address concerns that lay individuals may use both medical and/or cloth and paper masks incorrectly, masking techniques and norms need to be taught with targeted information to different demographics, just as proper handwashing and social distancing techniques have been taught.
\bibliographystyle{named}
\bibliography{refs}

\begin{thebibliography}{}

\bibitem[\protect\citeauthoryear{Abaluck \bgroup \em et al.\egroup
  }{2020}]{abaluckchevalierchristakisformankaplankovermund2020}
Jason Abaluck, Judith~A. Chevalier, Nicholas~A. Christakis, Howard~Paul Forman,
  Edward~H. Kaplan, Albert Ko, and Sten~H. Vermund.
\newblock The {Case} for {Universal} {Cloth} {Mask} {Adoption} and {Policies}
  to {Increase} {Supply} of {Medical} {Masks} for {Health} {Workers}.
\newblock {SSRN} {Scholarly} {Paper} ID 3567438, Social Science Research
  Network, Rochester, NY, April 2020.

\bibitem[\protect\citeauthoryear{{Académie nationale de
  médecine}}{2020}]{academienationaledemedecine2020}
{Académie nationale de médecine}.
\newblock Communiqué de l’{Académie} : "{Pandémie} de {Covid}-19 : mesures
  barrières renforcées pendant le confinement et en phase de sortie de
  confinement".
\newblock Technical report, Académie nationale de médecine, April 2020.

\bibitem[\protect\citeauthoryear{Bonabeau}{2002}]{bonabeau2002}
Eric Bonabeau.
\newblock Agent-based modeling: {Methods} and techniques for simulating human
  systems.
\newblock {\em Proceedings of the National Academy of Sciences of the United
  States of America}, 99(Suppl 3):7280--7287, May 2002.

\bibitem[\protect\citeauthoryear{Burch and Bunt}{2020}]{BurchBunt2020}
Jane Burch and Christopher Bunt.
\newblock Can physical interventions help reduce the spread of respiratory
  viruses?
\newblock {\em Cochrane Clinical Answers}, 2020.
\newblock Publisher: John Wiley \& Sons, Ltd.

\bibitem[\protect\citeauthoryear{Burgess and Horii}{2012}]{burgesshorii2012}
Adam Burgess and Mitsutoshi Horii.
\newblock Risk, ritual and health responsibilisation: Japan’s ‘safety
  blanket’ of surgical face mask-wearing.
\newblock {\em Sociology of Health and Illness}, 34(8):1184–1198, 2012.

\bibitem[\protect\citeauthoryear{Cheong}{2020}]{cheong2020}
Danson Cheong.
\newblock Coronavirus: Most workplaces to close, schools will move to full
  home-based learning from next week, says pm lee.
\newblock {\em The Straits Times}, Apr 2020.
\newblock
  https://www.straitstimes.com/singapore/health/most-workplaces-to-close-schools-will-move-to-full-home-based-learning-from-next.

\bibitem[\protect\citeauthoryear{Cowling \bgroup \em et al.\egroup
  }{2020}]{cowlingalingtsanglifongliaokwanleechiuwuwuleung2020}
Benjamin~J. Cowling, Sheikh~Taslim Ali, Tiffany W.~Y. Ng, Tim~K. Tsang, Julian
  C.~M. Li, Min~Whui Fong, Qiuyan Liao, Mike~YW Kwan, So~Lun Lee, Susan~S.
  Chiu, Joseph~T. Wu, Peng Wu, and Gabriel~M. Leung.
\newblock Impact assessment of non-pharmaceutical interventions against
  coronavirus disease 2019 and influenza in {Hong} {Kong}: an observational
  study.
\newblock {\em The Lancet Public Health}, 0(0), April 2020.
\newblock Publisher: Elsevier.

\bibitem[\protect\citeauthoryear{Davies \bgroup \em et al.\egroup
  }{2013}]{DaviesThompsonGiriKafatosWalkerBennett2013}
Anna Davies, Katy-Anne Thompson, Karthika Giri, George Kafatos, Jimmy Walker,
  and Allan Bennett.
\newblock Testing the efficacy of homemade masks: would they protect in an
  influenza pandemic?
\newblock {\em Disaster Medicine and Public Health Preparedness},
  7(4):413--418, August 2013.

\bibitem[\protect\citeauthoryear{{De Kai}}{2020}]{dekai2020}
{De Kai}.
\newblock The disastrous consequences of information disorder erupting around
  {COVID}-19: {AI} is preying upon our unconscious cognitive biases.
\newblock In {\em Boma {COVID-19} Summit}, March 2020.
\newblock \url{https://youtu.be/ZidC7oRd7Pc}, transcript at
  \url{http://dek.ai/unbias}.

\bibitem[\protect\citeauthoryear{Feng \bgroup \em et al.\egroup
  }{2020}]{FengShenXiaSongFanCowling2020}
Shuo Feng, Chen Shen, Nan Xia, Wei Song, Mengzhen Fan, and Benjamin~J. Cowling.
\newblock Rational use of face masks in the {COVID}-19 pandemic.
\newblock {\em The Lancet Respiratory Medicine}, 0(0), March 2020.
\newblock Publisher: Elsevier.

\bibitem[\protect\citeauthoryear{Fineberg}{2020}]{fineberg2020}
Harvey~V. Fineberg.
\newblock Ten {Weeks} to {Crush} the {Curve} {\textbar} {NEJM}.
\newblock {\em New England Journal of Medicine}, April 2020.

\bibitem[\protect\citeauthoryear{Ganyani \bgroup \em et al.\egroup
  }{2020}]{ganyanikremerchentornerifaeswallingahens2020}
Tapiwa Ganyani, Cécile Kremer, Dongxuan Chen, Andrea Torneri, Christel Faes,
  Jacco Wallinga, and Niel Hens.
\newblock Estimating the generation interval for {COVID}-19 based on symptom
  onset data.
\newblock {\em medRxiv}, 2020.

\bibitem[\protect\citeauthoryear{Gottlieb and
  Rivers}{2020}]{gottliebrivers2020_washpost20200303}
Scott Gottlieb and Caitlin~M. Rivers.
\newblock Quarantining cities isn’t needed. {But} a fast, coordinated
  response to covid-19 is essential.
\newblock {\em Washington Post}, March 2020.

\bibitem[\protect\citeauthoryear{Gottlieb \bgroup \em et al.\egroup
  }{2020}]{gottliebriversmcclellansilviswatson2020}
Scott Gottlieb, Caitlin Rivers, Mark~B. McClellan, Lauren Silvis, and Crystal
  Watson.
\newblock National coronavirus response: {A} road map to reopening.
\newblock Technical report, American Enterprise Institute (AEI0, March 2020.

\bibitem[\protect\citeauthoryear{Hansstein and
  Echegaray}{2018}]{hanssteinechegaray2018}
Francesca~Valeria Hansstein and Fabián Echegaray.
\newblock Exploring motivations behind pollution-mask use in a sample of young
  adults in urban china.
\newblock {\em Globalization and Health}, 14(1), Dec 2018.

\bibitem[\protect\citeauthoryear{Holland and Miller}{1991}]{hollandmiller1991}
John~H. Holland and John~H. Miller.
\newblock Artificial {Adaptive} {Agents} in {Economic} {Theory}.
\newblock {\em The American Economic Review}, 81(2):365--370, 1991.
\newblock Publisher: American Economic Association.

\bibitem[\protect\citeauthoryear{{Hong Kong Department of
  Health}}{2020}]{hkdoh2020}
{Hong Kong Department of Health}.
\newblock Latest recommendations by scientific committee on emerging and
  zoonotic diseases and scientific committee on infection control after
  reviewing cases of novel coronavirus infection.
\newblock Technical report, The Government of the Hong Kong Special
  Administration Region, Jan 2020.
\newblock Press Release, Scientific Committee on Emerging and Zoonotic Diseases
  and the Scientific Committee on Infection Control under the Centre for Health
  Protection (CHP) of the Department of Health.
  https://www.info.gov.hk/gia/general/202001/24/P2020012400762.htm.

\bibitem[\protect\citeauthoryear{Howard and {Fast.ai team}}{2020}]{howard2020}
Jeremy Howard and {Fast.ai team}.
\newblock Make and wear a homemade mask to slow the spread of {COVID}-19.
\newblock {\em \#Masks4All}, 2020.
\newblock https://masks4all.co.

\bibitem[\protect\citeauthoryear{Hunter \bgroup \em et al.\egroup
  }{2017}]{huntermacnameekelleher2017}
Elizabeth Hunter, Brian Mac~Namee, and John~D. Kelleher.
\newblock A {Taxonomy} for {Agent}-{Based} {Models} in {Human} {Infectious}
  {Disease} {Epidemiology}.
\newblock {\em Journal of Artificial Societies and Social Simulation}, 20(3):2,
  2017.

\bibitem[\protect\citeauthoryear{Hunter \bgroup \em et al.\egroup
  }{2018}]{huntermacnameekelleher2018}
Elizabeth Hunter, Brian Mac~Namee, and John Kelleher.
\newblock An open-data-driven agent-based model to simulate infectious disease
  outbreaks.
\newblock {\em PLOS ONE}, 13(12):e0208775, December 2018.

\bibitem[\protect\citeauthoryear{Jennings}{2020}]{jennings2020}
Ralph Jennings.
\newblock Not just coronavirus: Asians have worn face masks for decades.
\newblock {\em Voice of America}, Mar 2020.
\newblock
  https://www.voanews.com/science-health/coronavirus-outbreak/not-just-coronavirus-asians-have-worn-face-masks-decades.

\bibitem[\protect\citeauthoryear{Kuo}{2014}]{kuo2014}
Lily Kuo.
\newblock Chinese urbanites are spending millions on anti-pollution masks-and
  most of them don't do anything.
\newblock {\em Quartz}, Feb 2014.
\newblock
  https://qz.com/180830/chinese-urbanites-are-spending-millions-on-anti-pollution-masks-and-most-of-them-dont-do-anything/.

\bibitem[\protect\citeauthoryear{Leung \bgroup \em et al.\egroup
  }{2020}]{LeungChuShiuChanMcDevittHauYenLiIpPeirisSetoLeungMiltonCowling2020}
Nancy H.~L. Leung, Daniel K.~W. Chu, Eunice Y.~C. Shiu, Kwok-Hung Chan,
  James~J. McDevitt, Benien J.~P. Hau, Hui-Ling Yen, Yuguo Li, Dennis K.~M. Ip,
  J.~S.~Malik Peiris, Wing-Hong Seto, Gabriel~M. Leung, Donald~K. Milton, and
  Benjamin~J. Cowling.
\newblock Respiratory virus shedding in exhaled breath and efficacy of face
  masks.
\newblock {\em Nature Medicine}, pages 1--5, April 2020.
\newblock Publisher: Nature Publishing Group.

\bibitem[\protect\citeauthoryear{Leung}{2020}]{Leung2020_Time20200312}
Hillary Leung.
\newblock Why {Face} {Masks} {Are} {Encouraged} in {Asia}, but {Shunned} in the
  {U}.{S}.
\newblock {\em Time}, March 2020.

\bibitem[\protect\citeauthoryear{Li}{2014}]{li2014}
Grace Li.
\newblock China's face mask industry under scrutiny as pollution worsens.
\newblock {\em Reuters}, Mar 2014.
\newblock
  https://www.reuters.com/article/us-china-mask-pollution/chinas-face-mask-industry-under-scrutiny-as-pollution-worsens-idUSBREA2O0GI20140325.

\bibitem[\protect\citeauthoryear{Liu \bgroup \em et al.\egroup
  }{2020}]{LiuNingChenGuoLiuGaliSunDuanCaiWesterdahlLiuHoKanFuLan2020}
Yuan Liu, Zhi Ning, Yu~Chen, Ming Guo, Yingle Liu, Nirmal~Kumar Gali, Li~Sun,
  Yusen Duan, Jing Cai, Dane Westerdahl, Xinjin Liu, Kin-fai Ho, Haidong Kan,
  Qingyan Fu, and Ke~Lan.
\newblock Aerodynamic {Characteristics} and {RNA} {Concentration} of
  {SARS}-{CoV}-2 {Aerosol} in {Wuhan} {Hospitals} during {COVID}-19 {Outbreak}
  {\textbar} {bioRxiv}.
\newblock {\em bioRxiv}, March 2020.

\bibitem[\protect\citeauthoryear{MacIntyre \bgroup \em et al.\egroup
  }{2015}]{macintyresealedunghienngachughtairahmandwyerwang2015}
C.~Raina MacIntyre, Holly Seale, Tham~Chi Dung, Nguyen~Tran Hien, Phan~Thi Nga,
  Abrar~Ahmad Chughtai, Bayzidur Rahman, Dominic~E. Dwyer, and Quanyi Wang.
\newblock A cluster randomised trial of cloth masks compared with medical masks
  in healthcare workers.
\newblock {\em BMJ Open}, 5(4):e006577, April 2015.
\newblock Publisher: British Medical Journal Publishing Group Section:
  Infectious diseases.

\bibitem[\protect\citeauthoryear{Manjoo}{2020}]{manjoo2020_nytimes20200331}
Farhad Manjoo.
\newblock It’s {Time} to {Make} {Your} {Own} {Face} {Mask}.
\newblock {\em New York Times}, March 2020.

\bibitem[\protect\citeauthoryear{Morgunov \bgroup \em et al.\egroup
  }{2020}]{morgunovbaynomanawis2020}
Alexey Morgunov, Z.~Bayno, and R.~Manawis.
\newblock Status of face mask wearing around the world, March 2020.
\newblock
  https://docs.google.com/spreadsheets/d/1bzrUZe7qGGuYrHcYwyp-Ukh7tHW7NuZM\_rkdGACR9Dc.

\bibitem[\protect\citeauthoryear{Nebehay and
  Shalal}{2020}]{nebehayshalal2020_reuters20200403}
Stephanie Nebehay and Andrea Shalal.
\newblock {WHO} opens door to broader use of masks to limit spread of
  coronavirus.
\newblock {\em Reuters}, April 2020.

\bibitem[\protect\citeauthoryear{Russell and Norvig}{2009}]{russellnorvig2009}
Stuart Russell and Peter Norvig.
\newblock {\em Artificial {Intelligence}: {A} {Modern} {Approach}}.
\newblock Pearson, Upper Saddle River, 3 edition edition, December 2009.

\bibitem[\protect\citeauthoryear{Sande \bgroup \em et al.\egroup
  }{2008}]{SandeTeunisSabel2008}
Marianne van~der Sande, Peter Teunis, and Rob Sabel.
\newblock Professional and {Home}-{Made} {Face} {Masks} {Reduce} {Exposure} to
  {Respiratory} {Infections} among the {General} {Population}.
\newblock {\em PLOS ONE}, 3(7):e2618, July 2008.

\bibitem[\protect\citeauthoryear{Santarpia \bgroup \em et al.\egroup
  }{2020}]{SantarpiaRiveraHerreraMorwitzerCreagerSantarpiaCrownBrettmajorSchnaubeltBroadhurstLawlerReidLowe2020}
Joshua~L. Santarpia, Danielle~N. Rivera, Vicki Herrera, M.~Jane Morwitzer,
  Hannah Creager, George~W. Santarpia, Kevin~K. Crown, David Brett-Major,
  Elizabeth Schnaubelt, M.~Jana Broadhurst, James~V. Lawler, St.~Patrick Reid,
  and John~J. Lowe.
\newblock Transmission {Potential} of {SARS}-{CoV}-2 in {Viral} {Shedding}
  {Observed} at the {University} of {Nebraska} {Medical} {Center} {\textbar}
  {medRxiv}.
\newblock {\em medRxiv}, March 2020.

\bibitem[\protect\citeauthoryear{Service}{2020}]{Service2020}
Robert~F. Service.
\newblock You may be able to spread coronavirus just by breathing, new report
  finds.
\newblock {\em Science}, April 2020.

\bibitem[\protect\citeauthoryear{Servick}{2020}]{servick2020}
Kelly Servick.
\newblock Would everyone wearing face masks help us slow the pandemic?
\newblock {\em Science}, March 2020.

\bibitem[\protect\citeauthoryear{Tracht \bgroup \em et al.\egroup
  }{2010}]{TrachtValleHyman2010}
Samantha~M. Tracht, Sara Y.~Del Valle, and James~M. Hyman.
\newblock Mathematical {Modeling} of the {Effectiveness} of {Facemasks} in
  {Reducing} the {Spread} of {Novel} {Influenza} {A} ({H1N1}).
\newblock {\em PLOS ONE}, 5(2):e9018, February 2010.

\bibitem[\protect\citeauthoryear{Tracy \bgroup \em et al.\egroup
  }{2018}]{tracycerdakeyes2018}
Melissa Tracy, Magdalena Cerdá, and Katherine~M. Keyes.
\newblock Agent-{Based} {Modeling} in {Public} {Health}: {Current}
  {Applications} and {Future} {Directions}.
\newblock {\em Annual review of public health}, 39:77--94, April 2018.

\bibitem[\protect\citeauthoryear{Tufekci}{2020}]{tufekci2020_nytimes20200317}
Zeynep Tufekci.
\newblock Why {Telling} {People} {They} {Don}’t {Need} {Masks} {Backfired}.
\newblock {\em New York Times}, March 2020.
\newblock
  https://www.nytimes.com/2020/03/17/opinion/coronavirus-face-masks.html.

\bibitem[\protect\citeauthoryear{van Doremalen \bgroup \em et al.\egroup
  }{2020}]{VanDoremalenBushmakerMorrisHolbrookGambleWilliamsonTaminHarcourtThornburgGerberLloyd2020}
Neeltje van Doremalen, Trenton Bushmaker, Dylan~H. Morris, Myndi~G. Holbrook,
  Amandine Gamble, Brandi~N. Williamson, Azaibi Tamin, Jennifer~L. Harcourt,
  Natalie~J. Thornburg, Susan~I. Gerber, James~O. Lloyd-Smith, Emmie de~Wit,
  and Vincent~J. Munster.
\newblock Aerosol and {Surface} {Stability} of {SARS}-{CoV}-2 as {Compared}
  with {SARS}-{CoV}-1.
\newblock {\em New England Journal of Medicine}, 382(16):1564--1567, April
  2020.
\newblock Publisher: Massachusetts Medical Society \_eprint:
  https://doi.org/10.1056/NEJMc2004973.

\bibitem[\protect\citeauthoryear{{World Health Organization}}{2019}]{who2019}
{World Health Organization}.
\newblock Coronavirus disease ({COVID}-19) advice for the public: {When} and
  how to use masks. {World} {Health} {Organization}, 2019.
\newblock
  https://www.who.int/emergencies/diseases/novel-coronavirus-2019/advice-for-public/when-and-how-to-use-masks.

\bibitem[\protect\citeauthoryear{{World Health
  Organization}}{2020}]{who2020_20200406}
{World Health Organization}.
\newblock Advice on the use of masks in the context of {COVID}-19: interim
  guidance, 6 {April} 2020.
\newblock Technical Report WHO/2019-nCov/IPC\_Masks/2020.3, {World Health
  Organization}, April 2020.
\newblock Accepted: 2020-04-06T20:48:18Z Number:
  WHO/2019-nCov/IPC\_Masks/2020.3 Publisher: World Health Organization.

\bibitem[\protect\citeauthoryear{Yan \bgroup \em et al.\egroup
  }{2019}]{YanGuhaHariharanMyers2019}
Jing Yan, Suvajyoti Guha, Prasanna Hariharan, and Matthew Myers.
\newblock Modeling the {Effectiveness} of {Respiratory} {Protective} {Devices}
  in {Reducing} {Influenza} {Outbreak}.
\newblock {\em Risk Analysis}, 39(3):647--661, 2019.
\newblock \_eprint: https://onlinelibrary.wiley.com/doi/pdf/10.1111/risa.13181.

\bibitem[\protect\citeauthoryear{Yang}{2014}]{yang2014}
Jeff Yang.
\newblock A quick history of why asians wear surgical masks in public.
\newblock {\em Quartz}, Nov 2014.
\newblock
  https://qz.com/299003/a-quick-history-of-why-asians-wear-surgical-masks-in-public/.

\end{thebibliography}

\end{document}